**The Neoplasia as embryological phenomenon and its implication in the animal evolution and the origin of cancer. III. The role of flagellated cell fusion in the formation of the first animal in the Ediacaran ocean and evolutionary clues to the Warburg effect**

Jaime Cofre[1,*]

[1] Laboratório de Embriologia Molecular e Câncer, Federal University of Santa Catarina, room 313b, Florianópolis, SC, 88040-900, Brazil

[*] Corresponding author. Laboratório de Embriologia Molecular e Câncer, Universidade Federal de Santa Catarina, Sala 313b, Florianópolis, SC, 88040-900, Brazil

E-mail address: jaime.cofre@ufsc.br

**Abstract**

Cytasters have been underestimated in terms of their potential relevance to embryonic development and evolution. From the perspective discussed herein, structures such as the multiciliated cells of comb rows and balancers supporting mineralized statoliths and macrocilia in *Beroe ovata* point to a past event of multiflagellate fusion in the origin of metazoans. These structures, which are unique in evolutionary history, indicate that early animals handled basal bodies and their duplication in a manner consistent with a "developmental program" originated in the Ctenophora. Furthermore, the fact that centrosome amplification leads to spontaneous tumorigenesis suggests that the centrosome regulation process was co-opted into a neoplastic functional module. Multicilia, cilia, and flagella are deeply rooted in the evolution of animals and Neoplasia. The fusion of several flagellated microgametes into a cell with a subsequent phase of zygotic (haplontic) meiosis might have been at the origin of both animal evolution and the neoplastic process. In the Ediacaran ocean, we also encounter evolutionary links between the Warburg effect and Neoplasia.

**Phylogenetic records of ciliation and multiciliation**

The similarities between ctenophorean polyspermy and the arrangement of bilaterian cytasters have led to the suggestion of a polyspermic ancestor in the origin of animals (Salinas-Saavedra and Vargas, 2011). Cytasters (cytoplasmic asters) are centriole-based nucleation centers for microtubule polymerization that occur in large numbers in the cortical cytoplasm of oocytes and zygotes of bilaterian organisms (Kallenbach, 1985; Cantillana *et al.*, 2000). In protostomes and deuterostomes, cytasters are produced during oogenesis from vesicles of the nuclear membrane or the surrounding Golgi apparatus that move to the cortical cytoplasm (Kallenbach, 1985; Szöllosi and Ozil, 1991; Calarco, 2000). These structures associate with various cytoplasmic components and participate in the reorganization of the cortical cytoplasm after fertilization, patterning anteroposterior and dorsoventral body axes (Salinas-Saavedra and Vargas, 2011).

In the history of science, the earliest reported observations of cytasters seem to have been made by Thomas Morgan, according to a thorough review of the literature carried out by Naohidé Yatsu in 1905:

> "An experimental study of the cytasters was for the first time made by Morgan. In 1893 he saw refractile drops in the egg of *Arbacia* treated with the sea-water to which a little $NaCl_2$ (2 percent) had been added. In the winter of 1894-95 he extended his experiments on the egg of *Sphaerechinus* to see if the refractile drops, which he later found to be the cytasters, cause the division of cytoplasm." (Yatsu, 1905)

Yatsu himself analyzed enucleated egg fragments of *Cerebratulus lacteus* and characterized numerous cytasters, which he defined as asters or pseudasters (centriole-less asters) unconnected with nuclear matter. In doing so, he offered some valuable insights:

> "Whether or not an aster containing a centrosome (centriole) may arise in the egg-cytoplasm independent of preexisting centers is a cytological problem of high interest. This seemingly difficult question may be decided by a simple experiment. If we are able to produce an aster with the centriole in an egg-fragment containing no preexisting centrioles, we cannot escape the conclusion that these structures may be formed *de novo* in the egg-cytoplasm." (Yatsu, 1905)

The hypothesis raised by Naohidé Yatsu was correct: it is now known that centrioles can be formed in the absence of pre-existing centrioles through a process termed *de novo* centriole formation. The link between protists and the phylogenetic origin of multiciliation in animals is widely reported in the scientific literature (Allen, 1969; Machemer, 1972; Ruiz *et al.*, 1999; Stemm-Wolf *et al.*, 2005; Nikolaev *et al.*, 2006; Gogendeau *et al.*, 2011; Soh *et al.*, 2022). Interestingly, there is also strong evidence supporting that centrosomes in protists can be formed by a *de novo* mechanism. *Naegleria gruberi*, when under stress, rapidly differentiates into a flagellate, forming a complete cytoplasmic cytoskeleton, including two basal bodies and their flagella, via *de novo* biosynthesis (Fritz-Laylin *et al.*, 2010). A similar *de novo* mechanism has also been described in *Chlamydomonas* (Marshall, Vucica and Rosenbaum, 2001). Surprisingly, in multiciliated cells of the brain (Mercey *et al.*, 2019; Zhao *et al.*, 2019) and tracheal epithelial cells of mice (Nanjundappa *et al.*, 2019; Zhao *et al.*, 2019), pre-existing centrioles are not essential for the formation of deuterosomes (electron-dense ring-shaped structures peculiar to multiciliated cells). Such findings demonstrate the value of studying non-canonical (*de novo*) pathways of centriolar biogenesis and investigating similarities and divergences between these two biosynthetic systems (centriole-dependent and independent routes). Polo-like kinase 4 (Plk4) and SCL/TAL1 interrupting locus (Stil) proteins are essential components of both centriolar biogenesis pathways and are likely interconnected by the pericentriolar material (PCM) (Takumi and Kitagawa, 2022). This undoubtedly reaffirms the relevance of cytasters to the origin of animals in which geminin or multicilin have not yet been identified (e.g., ctenophores) (Meunier and Azimzadeh, 2016).

The interest in understanding the control of centrosome/centriole amplification lies also in its relationship with tumorigenesis (Nigg, 2006; Nigg and Raff, 2009; Chan, 2011; Godinho and Pellman, 2014; Levine *et al.*, 2017). Plk4 overexpression in a mouse model of intestinal neoplasm resulted in spontaneous tumorigenesis (Levine *et al.*, 2017). The Stil protein, on the other hand, functions as an oncogenic factor that controls cell cycle in a cilium-dependent manner (Li *et al.*, 2022), acting in a modular fashion, which allows to speculate on the co-opting of ciliogenesis and multiciliogenesis into a neoplastic functional module (NFM) at the beginning of animal evolution. Centrosome amplification is a characteristic feature of cancer. Boveri's hypothesis formulated more than 100 years ago indicated that centrosome amplification promoted tumorigenesis (Boveri, 1929). As I will show in this article, the fusion of multiflagellate cells might have evolved into a form of Neoplasia that functioned as an evolutionary engine. I must also recognize the importance of centriole-less centrosomes (Meraldi and Nigg, 2002; Manandhar, Schatten and Sutovsky, 2005; Schatten, 2008; O∅Connell, 2021) and centriolar satellites (Odabasi, Batman and Firat-Karalar, 2020; Prosser and Pelletier, 2020) and their profound implications for experimental biology. In particular, the observation of an acentriolar PCM in neurons (O∅Connell, 2021) led me to think about an evolutionary approach founded on Neoplasia— a topic I will explore in another article.

Finally, another observation that cannot be treated as inconsequential for experimental biology, evolutionary theory, and oncology is perhaps one of the most relevant characteristics of ctenophores: their multicilia. Comb rows, a prominent example of compound cilia (i.e., cilia formed by several multiciliated cells) are known to occur only and certainly in ctenophores (Nielsen, 1987). In *Beroe ovata*, macrocilia consist of cylindrical protrusions with several hundreds of parallel axonemes surrounded by a plasma membrane, forming a structure that falls outside the conventional definitions of cilia (Nielsen, 1987; S. L. Tamm and Tamm, 1988). These multiciliated cells have numerous basal bodies (Jokura *et al.*, 2022) and are capable of regenerating (Tamm, 2012) through a process dependent on cell proliferation (Ramon-Mateu *et al.*, 2019), similar to that seen in the cnidarian *Nematostella* (DuBuc, Traylor-Knowles and Martindale, 2014). These observations demonstrate that multiciliary complexity reached an apex in ctenophores, decreasing thereafter in the course of animal evolution. In vertebrates, there is a so-called "developmental program" of cilia involving the proteins geminin and multicilin (Stubbs *et al.*, 2012; Zhou *et al.*, 2020). This topic has been little studied in invertebrate animals. On the other hand, the Notch protein is known to be absent in ctenophores. A better understanding of the developmental pathways of *de novo* centriole formation may help elucidate the role of cytasters in animal evolution, as well as their relation to protists. It is also important to note that phylogenetic studies place unicellular Holozoa at the base of animal evolution (Suga *et al.*, 2013; Grau-Bové *et al.*, 2017; Hehenberger *et al.*, 2017), and these organisms do not have multicilia, but flagella (Hehenberger *et al.*, 2017; Mylnikov *et al.*, 2019). On the basis of such observations, it is possible to rule out the occurrence of a fusion event with *Paramecium*-like features at the beginning of animal evolution and focus instead on the relevance of a fusion process involving multiple flagellated cells, which supports my proposal. Multiflagellate fusion attained such relevance by combining physical and chemical forces in the cellular context of the Ediacaran ocean. Next, we will see how the preponderance of the actin cytoskeleton in meiosis, cell fusion, and ciliogenesis was decisive to the formulation of my hypothesis.

**The role of calcium and actin in multiflagellate fusion**

Cytasters are absent in ctenophore oocytes, as confirmed by immunofluorescence and microtubule polymerization techniques (Houliston *et al.*, 1993). In *B. ovata*, polyspermy produces massive incorporation of asters, considerably expanding the microtubular cytoskeleton of the fertilized oocyte (Houliston *et al.*, 1993). Cortical cytoplasmic movements occurring after fertilization also closely resemble those of the bilaterian ectoplasm. I also understand that one of the central aspects of my proposal of multiflagellate fusion as the initial event leading to the genesis of metazoans is that we (animals) are linked by a flagellar proteome highly conserved throughout evolution (Merchant *et al.*, 2007). Polycystic kidney disease (in humans and mice) is triggered by a homolog of an intraflagellar transport protein of *Chlamydomonas* (Kozminski *et al.*, 1993; Pazour *et al.*, 2000). That is, in this disease, there is a defect in the formation of a ciliary/flagellar structure.

Another interesting aspect is the complexity of the flagellum proteome, which contains approximately 650 proteins (high to moderate confidence) (Pazour *et al.*, 2005; Davis, Brueckner and Katsanis, 2006), including a large number of motors, signal transducers, and disease-linked proteins. Thus, ciliopathies guide us to a flagellar past at the origin of metazoans. Proteins of the flagellum and its assembly system (intraflagellar transport) have played an inordinately important role in organogenesis (Lindemann, 2022). The large number of organs whose functions depend on cilia demonstrate the extent to which these structures are impregnated throughout embryogenesis (Eliasson *et al.*, 1977; Afzelius, 1995; Essner *et al.*, 2002; Badano *et al.*, 2006; Fliegauf, Benzing and Omran, 2007; Marshall, 2008; Sharma, Berbari and Yoder, 2008; Schou, Pedersen and Christensen, 2015; Hilgendorf, Johnson and Jackson, 2016; Pigino, 2021). The article entitled "The sperm, a neuron with a tail" is symptomatic in that it shows how flagellar proteins are used to assemble neurons (Meizel, 2004). Flagellar proteins also participate in neuronal migration in the cerebral cortex during embryogenesis (Morris *et al.*, 1998), in the release of synaptic vesicles (Li *et al.*, 2017), and in alterations such as hydrocephalus (Davy, 2003).

The origin of flagella as chemotaxis-sensitive (Yoshida and Yoshida, 2011) sensory structures (Marshall and Nonaka, 2006; Singla and Reiter, 2006) recruited in the context of embryonic development is also fundamental to the framework of my hypothesis. Ctenophores are extremely mobile mechanosensory animals highly dependent on ciliogenesis (Tamm, 2014). It is likely that flagellar proteins contributed significantly to the formation and evolution of metazoan sensory organs. In *Chlamydomonas*, action potentials are detected exclusively by the flagellum (Harz and Hegemann, 1991; Harz, Nonnengasser and Hegemann, 1992). Moreover, the ciliary distribution of voltage-sensitive sodium and calcium channels suggests that action potentials evolved in the flagellum in most protists (Dunlap, 1977; Machemer and Ogura, 1979; Fujiu *et al.*, 2009). An excellent hypothesis has already been proposed in an attempt to explain how action potentials left the ciliary context and expanded to neuronal and muscle cells, giving rise to contractile mechanosensory cells (Brunet and Arendt, 2016). From the point of view of my hypothesis, this expansion only occurred in an embryonic context after recruitment (co-option) of flagellar proteins into an essentially proliferative (Li *et al.*, 2022), mechanical (Ishikawa and Marshall, 2014), and physical module intimately linked to Neoplasia, as will be discussed below. The expansion of action potentials to neuronal cells also occurred in a physical and embryonic context related to neuromesodermal stem cells.

Other clues to the recruitment of cilia/flagella into an NFM can be found in the Wnt and Hedgehog (Hh) signaling pathways. Cilia, through the Jouberin protein, specifically regulate the Wnt pathway by limiting the nuclear entry of -catenin. Unciliated cells have potentiated Wnt responses, whereas cells with multiple cilia have inhibited responses (Lancaster, Schroth and Gleeson, 2011). Localization studies of Wnt in ctenophores (Jager *et al.*, 2013) revealed expression around the mouth margin at the oral pole (where macrociliary cells are coincidently located) (S. Tamm and Tamm, 1988), in locomotor ciliated comb rows (neurosensory structures), and at sites containing proliferative stem cells. The relationship of Wnt pathways with cancer is well established (Lai, Chien and Moon, 2009) and explains the predisposition of cancer cells to lose cilia, as will be described later in this article. As for Hh signaling, in mice, it requires the intraflagellar transport proteins Wim, Polaris, and Kif3a (Huangfu *et al.*, 2003). A systematic study identified components of the Hh pathway associated with cilia in invertebrates, indicating a long evolutionary history and providing evidence that, in *Drosophila*, the relationship between Hh and cilia was lost, but Fu and Cos2 (ancestral flagellar proteins) were incorporated as integral components of the Hh pathway (Roy, 2012). It could not be otherwise, given that the Hh pathway depends on ciliary assembly and aberrant Hh signaling is associated with human cancer (Han *et al.*, 2009; Wong *et al.*, 2009).

For the purposes of my hypothesis, the cell in meiosis (haplont) originated from the fusion of multiple flagellated cells at the beginning of metazoan evolution will be referred to as a multiflagellate cell (Figure 1). Considering that flagella are sensory structures and protists possess a wide network of nuclear actin (Soyer-Gobillard, Ausseil and Géraud, 1996; Berdieva *et al.*, 2016), it is possible to infer that this early multiflagellate cell must have had incredible physical and excitatory differences in the plasma membrane, with a wide arrangement of calcium and sodium channels (Dunlap, 1977; Machemer and Ogura, 1979; Fujiu *et al.*, 2009) and a cytoplasm capable of propagating action potentials (Hunley, Uribe and Marucho, 2018) and transmitting signals, conducted by F-actin, from the plasma membrane to the nucleus (Frieden and Gatenby, 2019) (Figure 2). An example of syncytia that propagate action potentials is found in the fungus *Neurospora crassa*, where they produce long-lasting potentials and considerable increases in membrane conductance (Slayman, Scott Long and Gradmann, 1976). Another example is seen in the sponge *Rhabdocalyptus dawsoni* (Hexactinellida), wherein conduction of electrical activity takes place in the trabecular syncytium (Lawn, Mackie and Silver, 1981; Leys and Mackie, 1997) and penetrates into flagellated chambers where the effector response is generated (Mackie, Lawn and De Ceccatty, 1983).

Nevertheless, it must also be considered that the presence of cilia/flagella would be incompatible with the division of the first zygote. Today we understand that centrosomes, cilia, and cell cycle are closely associated in a modular way (Tucker, Pardee and Fujiwara, 1979; Tucker, Scher and Stiles, 1979). Therefore, I speculate that it was first necessary to occur the recruitment of cilia/flagella into the NFM. There is already evidence that the ciliary protein Stil controls the cell cycle and functions as an oncogenic engine (Li *et al.*, 2022), as there is evidence that ciliary control reaches directly into the nucleus. Ciliary proteins are important components in DNA replication and repair pathways (Attanasio, 2015) responsible for preserving genomic integrity. Overexpression of Aurora kinases leads to centrosome amplification, chromosomal instability, increased mitotic errors, and susceptibility to carcinogenesis (Meraldi, Honda and Nigg, 2004). Also, the architecture of the actin cytoskeleton must have been impregnated in the multiflagellate cell, and mechanical aspects were relevant at the beginning of multicellularity. The only controversial points refer to the role of the actin cytoskeleton in the syngamy of ctenophores.

**Controversies over the role of actin in meiosis and syngamy**

Actin microfilaments are found in dynamic undulating membrane folds that extend from the cell cortex to the entire surface of the oocyte (Houliston *et al.*, 1996). *B. ovata* undergoes cortical reorganization around each sperm nucleus, producing a local tuft of long microvilli in response to sperm entry (Carré, Rouvière and Sardet, 1991). There is also local exocytosis and accumulation of mitochondria and autofluorescent vesicles recruited from the surrounding cortical region. Microtubules tend to run parallel to the plasma membrane, interacting closely with layers of endoplasmic reticulum (Houliston *et al.*, 1993). Strangely, the role of microtubules was preponderant in meiosis and syngamy (Houliston *et al.*, 1993; Rouvière *et al.*, 1994). It seems as if there is a void in evolution. The actin cytoskeleton is predominant in unicellular holozoans (Mylnikov *et al.*, 2019; Phillips *et al.*, 2021) and extremely important, for example, in the meiosis of cnidarians (Amiel and Houliston, 2009) and ascidians (Prodon, Sardet and Nishida, 2008) but no longer in that of ctenophores. What could have happened? Why does F-actin not seem to have a relevant role in *B. ovata* meiosis?

Assays using cytochalasin B might help understand this issue. Evelin Houliston noted in her studies on *B. ovata* that oocytes treated with cytochalasin B often display abnormal and erratic protrusions, as well as abnormal contractions on the cell surface, making it difficult to assess the normal function of actin filaments (Houliston *et al.*, 1993). Interestingly, these protrusions or bulgings are very similar to the bleb-like protrusions resulting from altered actin cytoskeleton reorganization induced by Yki loss in *Capsaspora owczarzaki* (Phillips *et al.*, 2021). *Capsaspora* is an amoeboid organism characterized by actin-rich filopodial projections and F-actin-sustained locomotion. Also in the cnidarian *Clytia hemisphaerica,* the role of actin microfilaments during oocyte growth could not be evaluated because treatments of the gonads with cytochalasin B or latrunculin B caused rapid tissue dissociation (Amiel and Houliston, 2009). Studies showed that treatment of neuronal growth cones with cytochalasin B produced side effects. Microtubules are restricted to the axis of neurites and a central domain of growth cones. Within approximately 5 min of cytochalasin B exposure, microtubules began to extend into the lamellar region, often reaching the peripheral margin.

After removal of cytochalasin B, microtubules were restored to their former central location (Forscher and Smith, 1988). Therefore, the results need to be interpreted with great caution, as these agents cause global and extreme changes to the actin cytoskeleton, rather than small localized changes (Smith, Lake and Johnson, 2020). With the use of cytochalasin B in a polyspermic condition (1 to 30 spermatozoa), microtubules would likely expand throughout the cell. It is thus possible to conclude that there are consistent studies on the role of microtubules in *B. ovata*, but there is limited information on actin filaments.

Ciliogenesis is another strong argument supporting the need to re-examine the actin cytoskeleton of ctenophores. These organisms have the largest-known ciliary structures and exploit the mobile and sensory functions of cilia in most of their behaviors (Tamm, 2014). There is evidence of a photoreceptor function of cilia with a 9+0 pattern (characteristic of primary cilia) (Horridge, 1964) and also of pressure-receptor cilia (Hernandez-Nicaise, 1984). In this regard, studies on the differentiation of macrociliary cells in *B. ovata* labia show the important role of actin filaments in the transport and migration of basal bodies to the apical surface (S. Tamm and Tamm, 1988), underscoring the need for re-examination of the actin cytoskeleton. The ciliary beating of comb rows for locomotion is one of the most important characteristics of ctenophores (Inoué, 1959). Close associations between cilia and the actin cytoskeleton in multiciliary cells are evident from the beginning of ciliogenesis, as seen in the participation of the actin cytoskeleton for centrosome/basal body migration and for the anchoring of cilia to the actin cytoskeleton by focal adhesion complexes (Smith, Lake and Johnson, 2020; Zhao, Khan and Westlake, 2022). This interdependence between cilia and F-actin has been characterized in vertebrates, revealing a surprising degree of conservation throughout evolution (Sorokin, 1968; Boisvieux-Ulrich, Laine and Sandoz, 1985).

On the other hand, the incredible similarity in ciliogenesis between ctenophores, mammalian lung cells, and quail cells shows that cilia can form *de novo* and by an acentriolar pathway (Sorokin, 1968; S. Tamm and Tamm, 1988; Boisvieux-Ulrich, Lainé and Sandoz, 1990). Stabilization of the actin cytoskeleton by RhoA has been recognized as a dominant inducer of ciliogenesis (Pan *et al.*, 2007), and treatment with cytochalasin D results in inactivation of the transcriptional coactivators YAP and TAZ of the Hippo pathway (J. Kim *et al.*, 2015). Recent studies have shown that mechanical stimuli such as cell shape and extracellular matrix contractility and rigidity are the main determinants of ciliogenic activity (Pitaval *et al.*, 2010; Ishikawa and Marshall, 2014). Furthermore, a signaling pathway involving RhoA, YAP, and Myc identified in polycystic kidney disease is dependent on intraflagellar transport proteins (Cai *et al.*, 2018).

In relation to Neoplasia, numerous studies have shown that repression of the formation of primary cilia is closely related to carcinogenesis, once again underscoring the importance of re-examining the contribution of the actin cytoskeleton to these processes (Seeley *et al.*, 2009; Basten *et al.*, 2013; Nobutani *et al.*, 2014; Du *et al.*, 2018; Qie *et al.*, 2020). The Stil protein would be involved in cell cycle regulation through primary cilia in tumor cells (Li *et al.*, 2022). Also, a recent study showed what could be the origin of the close relationship between actin and cell proliferation. The Hippo (YAP/TAZ/Yorkie) routes of *C. owczarzaki* control actin cytoskeleton dynamics and multicellular morphogenesis (Phillips *et al.*, 2021). Co-option of Hippo pathways for proliferation control is situated at the very beginning of metazoan evolution, and Hippo pathway components are enriched in proliferative tissues of ctenophores and cnidarians (Coste *et al.*, 2016). The only exception is the absence of Yorkie in ctenophores. Thus, it was hypothesized that co-option of the Hippo pathway for the control of cell proliferation in early metazoans might have provided a convenient mechanism linking cell proliferation to the cytoskeletal and mechanical properties of cells (Phillips *et al.*, 2021). I also speculate that a highly stressful multicellular phenomenon during multiflagellate fusion might have co-opted the control of ciliogenesis into an essentially proliferative NFM (Figure 2). Current evidence indicates that ciliogenesis is controlled by RhoA (Pan *et al.*, 2007) and YAP/TAZ (J. Kim *et al.*, 2015), and is probably related to Myc (Cai *et al.*, 2018) and cell proliferation (Li *et al.*, 2022).

Finally, an evolutionary approach points again to a re-examination of the important role of the actin cytoskeleton in ctenophore meiosis. In *C. hemisphaerica*, exaggerated cortical contractions were observed across oocytes in the period between germinal vesicle breakdown (GVBD) and polar body emission. These contractions are actin-dependent (Amiel and Houliston, 2009). In ascidians, actin-dependent contractions are important for spindle localization and polar body formation (Mori *et al.*, 2011), and the contractions of a nuclear network of actin are crucial to delivering chromosomes to the microtubule spindle (Lénárt *et al.*, 2005). In ctenophores, filament-dependent actin contraction appears to participate in surface movements of the zygote before the first cleavage. Cytochalasin B induced abnormal contractions (Houliston *et al.*, 1996). On the other hand, studies on RNA localization during meiosis confirmed one microtubule-dependent and two microtubule-independent mechanisms in cnidarians (Amiel and Houliston, 2009). In ascidians, RNA localization mechanisms are completely actin-dependent (Prodon, Sardet and Nishida, 2008). These findings suggest a tendency toward microtubule-independent mechanisms at the beginning of evolution, but the important role of actin is always preserved. A study re-examining the role of F-actin in *Xenopus laevis* showed that meiotic spindle anchoring and rotation in the oocyte cortex are F-actin-dependent (Gard, Cha and Roeder, 1995) and that microtubules help remodel the actomyosin network (Waterman-Storer *et al.*, 2000). Finally, participation of F-actin in the positioning of the meiotic spindle represents an ancient and well-conserved function in evolution (Amiel *et al.*, 2009); therefore, it is not possible to imagine a lack of association between the actin cytoskeleton and *B. ovata* meiosis.

In the *B. ovata*, polyspermic fertilization (Carré, Rouvière and Sardet, 1991) occurs immediately before or during the formation of the first polar body (Carré and Sardet, 1984). After several spermatozoa enter the oocyte, cytoplasmic components become associated with the supernumerary sperm nuclei, each of which forms a spherical region called the sperm pronuclear zone (SPZ). The SPZ contains cortical granules, mitochondria, endoplasmic reticulum, and other cytoplasmic components, including the nuclear envelope of the sperm pronucleus. Each SPZ is organized by the centriole associated with each male pronucleus (Carré and Sardet, 1984). The formation of this new cortical cytoplasmic configuration is driven by microtubule-mediated waves and is of utmost importance for the establishment of the oral–aboral axis (Houliston *et al.*, 1993), which occurs after fusion of pronuclei and the first cleavage cycle (Carré and Sardet, 1984; Houliston *et al.*, 1993). As cell development and division progress, the nuclear envelopes of supernumerary male pronuclei are ruptured, their DNA is degraded, and their giant asters shrink until they disappear, presumably becoming integrated into the microtubule network (Houliston *et al.*, 1993).

**Calcium in cell fusion**

Within the framework of my hypothesis, multiflagellate fusion is understood as a physical and biological process responsible for the recruitment of all elements that triggered NFM formation (Figure 2). In this section, I will first demonstrate the importance of calcium in multiflagellate fusion events, which allowed structural consolidation of the actin cytoskeleton within the cell nucleus, thereby providing the conditions for the rupture of the nuclear lamina, mechanosensory mechanisms, chromatin organization, and DNA replication, among other important events for NFM co-opting. It is undeniable that the beginning of animal life is linked to the chemistry of the $Ca^{2+}$ ion. Increased $Ca^{2+}$ levels in specific niches of the marine environment changed the fate of living forms inhabiting the planet. Furthermore, it can be affirmed without a shadow of a doubt that $Ca^{2+}$, through its "fusogenic" properties, creates mechanical asymmetries and facilitates the formation of the fusion pore (Tahir *et al.*, 2020), modifying the physical characteristics of life. As observed in a previous study:

> "Because of the central role of $Ca^{2+}$ in regulating biomembrane fusion, we cultured mated cells in a $Ca^{2+}$-deficient medium which prevented sexual development. The subsequent addition of $CaCl_2$ not only speeded up the events of early sexual development, but also induced extensive fusions which were reflected in the formation of extremely large multinucleate syncytia." (McConachie and O'Day, 1986)

Other studies point out that both cell fusion and zygote nuclear differentiation (swelling, migration, and nuclear fusion) are $Ca^{2+}$-dependent events in *Dictyostelium discoideum* (Szabo, O'Day and Chagla, 1982). Protists such as *Scenedesmus quadricauda* (phylum Chlorophyta) (Margulis *et al.*, 1989) also exhibit $Ca^{2+}$-dependent cell fusion, forming spiny colonies (Kylin and Das, 1967; Trainor, 1969). Even in vertebrates such as chicken, erythrocyte fusion is inhibited by EDTA (Ahkong *et al.*, 1973) or EGTA (Volsky and Loyter, 1978), showing that calcium is a general requirement for membrane fusion. Here, it is important to highlight that syncytium formation and multiflagellate fusion are thermodynamically favorable biophysical phenomena that, nonetheless, need to overcome several energy barriers (Hernández and Podbilewicz, 2017), involving mechanical forces and lipid rearrangements that must be effected by specialized molecules.

Biological fusogens overcome at least four energy barriers: (i) approximation of biological membranes (cadherins/$Ca^{2+}$ may contribute to this), (ii) dehydration of plasma membranes by apposition, (iii) fusion of outer monolayers or hemifusion with nucleation of a stalk, and (iv) opening and expansion of the fusion pore (or pores, in the case of polyspermy). Pore expansion can be accelerated by excess $Ca^{2+}$ (Tahir *et al.*, 2020). Fusogenic molecules have been extensively characterized in protists, unicellular holozoans, basal metazoans, and other animals. For example, HAP2 was detected in *Chlamydomonas reinhardtii*; soluble *N*-ethylmaleimide-sensitive factor-attachment protein receptor (SNARE) complexes were identified in *C. reinhardtii*, *D. discoideum* (Sanderfoot, 2007; Kienle, Kloepper and Fasshauer, 2009), and in the unicellular holozoans *C. owczarzaki*, *Monosiga brevicollis*, and *Salpingoeca rosetta* (Burkhardt, 2015); fusion family proteins were found in ctenophores (e.g., *Pleurobrachia pileus*) and arthropods (Avinoam *et al.*, 2011); and AFF-1 and EFF-1 were detected in *Caenorhabditis elegans* (Mohler *et al.*, 2002; Sapir *et al.*, 2007). These observations indicate how $Ca^{2+}$, together with fusogens, would participate in the sexual cycle (Brukman, Li and Podbilewicz, 2021) and life cycle of protists (Demin, Berdieva and Goodkov, 2022) and opisthokonts (Mela, Rico-Ramírez and Glass, 2020), and allow envisaging an Ediacaran ocean with multiple multinucleated syncytial forms.

Another important aspect is the intracellular control of calcium concentration, which must have been decisive for the use of the physical properties of viscoelasticity during multiflagellate fusion (Taylor *et al.*, 1973). At low concentrations ($<10^{-7}$ M), $Ca^{2+}$ dissociates actin microfilaments. At concentrations close to $10^{-7}$ M, $Ca^{2+}$ seems to stabilize them, whereas, at higher concentrations, the ion induces, in the presence of ATP, the formation of microfilament groups and initiates a process similar to contraction (Taylor *et al.*, 1973). Thus, an increase in $Ca^{2+}$ levels in the Ediacaran ocean might have helped to establish a wide network of actin cytoskeleton with contractile capacity throughout cells as well as a connection of the cytoskeleton with fusion proteins. Actin polymerization appears to be involved in myoblast fusion in *Drosophila* (Massarwa *et al.*, 2007; Haralalka *et al.*, 2011; Kim *et al.*, 2015; Deng, Azevedo and Baylies, 2017) and in the epidermis of *C. elegans* (Yang *et al.*, 2017), always placing membranes closer to each other (Yang *et al.*, 2017), generating transient communication structures between cells (fusogenic synapse) (Rout,

Preußner and Önel, 2022), and producing mechanical stress that is purported to be fundamental for myoblast fusion (Kim *et al.*, 2015).

Recent discoveries surprisingly show the existence of nuclear actin polymers (Mori *et al.*, 2014; Okuno *et al.*, 2020; Wesolowska *et al.*, 2020) and explain how a physical pressure gradient mediated by F-actin is able to modulate the oocyte transcriptome, which is essential for embryonic development (Almonacid *et al.*, 2019). Studies on *B. ovata* revealed a wide actin network extending from the nuclear periphery to the apical surface of macrociliary cells (S. Tamm and Tamm, 1988), whose mechanosensory activity has been demonstrated (Tamm, 1983). This mechanotransduction role of actin reinforces its well-known functions as a transcriptional regulator, a constitutive component of RNA polymerases, and an essential factor for chromatin remodeling complexes, as well as its participation in the formation of heterogeneous nuclear ribonucleoprotein complexes, gene activation (Miralles and Visa, 2006; Zheng *et al.*, 2009; Yamazaki, Yamamoto and Harata, 2015), and even chromatin organization (Baarlink *et al.*, 2017; Mogessie and Schuh, 2017; Parisis *et al.*, 2017; Scheffler *et al.*, 2022). All these functions lend support to a biophysical phenomenon, linked to actin, with possible implications for NFM recruitment.

Multiflagellate fusion also prompts us to think about the importance of cell fusion phenomena, how they are impregnated in embryonic development, and the hypothesis that these phenomena were actually co-opted to the NFM. The cell fusion process is very important during embryonic development, for instance, for the formation of skeletal musculature (Yaffe and Feldman, 1965), osteoclasts (Solari *et al.*, 1995; Vignery, 2000), liver cells (Lizier *et al.*, 2018), and human trophoblasts (Benirschke, 1998; Huppertz, Bartz and Kokozidou, 2006; Huppertz and Borges, 2008). However, it is in studies of *C. elegans* that it is possible to find the decisive elements for understanding the co-option of cell fusion to NFM. *C. elegans* is an excellent model system of embryonic development (Corsi, Wightman and Chalfie, 2015), given that the lineages of all somatic nuclei of the adult hermaphrodite are widely known, and their development process is essentially invariant (Podbilewicz and White, 1994). About a third of the cells produced by this organism are subsequently fused into 44 syncytia in a highly reproducible and stereotyped fashion (Shemer and Podbilewicz, 2000). Moreover, analyses of cell fusion for hypodermis formation showed an initial migratory process exhibiting an ostensible rearrangement of adherent junctions prior to cell fusion (Podbilewicz and White, 1994). These pieces of evidence are very suggestive and led me to speculate on the inclusion of cell fusion pathways in Neoplasia, because of its modular characteristic.

This manner of thinking of cell fusion as a mechanism that has been co-opted to the NFM allows us to understand why spontaneous cell reprogramming and regeneration occur upon cell fusion (Terada *et al.*, 2002; Vassilopoulos and Russell, 2003; Sullivan and Eggan, 2006; Alvarez-Dolado, 2007; Lluis and Cosma, 2009a; Manohar and Lagasse, 2009; Soza-Ried and Fisher, 2012) and why the reprogramming efficiency increases after cell fusion via activation of genes involved in early embryogenesis, such as Nanog, Wnt/ -catenin, Sall4, MAPK/ERK, and Akt genes (Silva *et al.*, 2006; Lluis *et al.*, 2008; Nakamura *et al.*, 2008; Wong *et al.*, 2008; Lluis and Cosma, 2009b). Also, the hypothesis of cell fusion as co-opted to Neoplasia would allow us to correct the wrong conception that germ cells, which fuse during fertilization, would have a lower reprogramming potential (commonly termed pluripotency) or would be exclusively fated to produce specific cell types because they are terminally differentiated (Lluis and Cosma, 2009a). Fertilization is a physiological and physical phenomenon producing a totipotent zygote that subsequently undergoes intense reprogramming to allow cell differentiation for embryo formation. Therefore, it can be understood that germ cells differentiate to produce undifferentiation (totipotency) after fertilization and that such a potency reveals itself during morphogenesis.

As can be seen, my hypothesis of co-option of cell fusion to Neoplasia predicts the propensity of tumor cells to spontaneously fuse in culture (Wakeling, 1994), a phenomenon that has been known for several decades but remains unexplained and for which there was little interest in investigating (Duelli and Lazebnik, 2003). In 1911, Otto Aichel proposed that malignancy could be a consequence of hybridization between leukocytes and somatic cells and that differences between cells within tumors could be explained by an uneven distribution of chromosomes (for a review of the topic, see Rachkovsky *et al.*, 1998). Fusibility is not limited to a particular cell or tumor type, and fusion events can occur between tumor and non-tumor cells (Larizza, Schirrmacher and Pflüger, 1984), ultimately accentuating metastatic characteristics (Miller and Ruddle, 1976; Rachkovsky *et al.*, 1998). It should also be emphasized that tumor cell fusion is not limited to tissue culture conditions. Human tumor cells injected into hamsters produced metastases composed of cells from both animal species (Goldenberg, Pavia and Tsao, 1974).

In agreement with my hypothesis, there is consensus that fusion between somatic cells (stem cells or not, an event mimicking fertilization) produces hybrids that, after a proliferation period (which mimics the initial embryo formation phase), lead to two possible conditions: (i) differentiation/regeneration or (ii) cancer (figure 3 in Bjerkvig *et al.*, 2005; figure 1 in Lluis and Cosma, 2009a; Manohar and Lagasse, 2009; Goding, Pei and Lu, 2014; Demin, Berdieva and Goodkov, 2022). Some authors attribute cancer to genetic instability after hybrid formation (Bjerkvig *et al.*, 2005), but this claim can no longer be sustained (Vitale *et al.*, 2011). As we will see later in this article, polyploidy is a physiological and eminently evolutionary phenomenon of high biocomplexity (Fox *et al.*, 2020a; Van de Peer *et al.*, 2021; Anatskaya and Vinogradov, 2022). Finally, a perfect example of the two faces of Neoplasia is given by the ability of normal human stem cells originating from a grafted kidney to migrate to the skin, fuse with skin cells to adopt a keratinocyte phenotype, and produce a skin carcinoma (Aractingi *et al.*, 2005). Another interesting observation: it is not uncommon for some types of cancers to be associated with cell fusion molecules such as syncytin (Bjerregaard *et al.*, 2006; Larsson *et al.*, 2007).

In this new context of the article, we must inevitably discuss evidence supporting the somatic origin of teratocarcinomas (Mintz, Cronmiller and Custer, 1978), which reinforces the concept that embryonic somatic cells harbor all chromosomes and, potentially, all genes (albeit some might be turned off, depending on differentiation) that would allow the NFM to be activated under specific conditions. Also, normal cells can become cancerous, both *in vitro* and *in vivo*, after introduction of an adequate combination of oncogenes (Ilyas *et al.*, 1999; Bachoo *et al.*, 2002). Co-transfection of different combinations of oncogenes results in different phenotypic effects. What is missing in somatic cell fusion experiments to improve the efficiency of cell reprogramming? If we know all genes and chromosomes found in the metazoan genome, what is lacking to achieve complete NFM activation, in correct order, according to the exaptations proposed in this hypothesis?

It goes without saying that the more conserved the normal structure of the initial phase of embryogenesis, the greater the odds of success in forming the embryo (Gurdon, Laskey and Reeves, 1975; Wilmut *et al.*, 1997; Wakayama *et al.*, 1998). Also without doubt, physical forces are decisive for the embryonic process. For example, a mechanical biomembrane model has been developed to explain the behavior of the zona pellucida after fertilization (Yu Sun *et al.*, 2003). This may have important implications for intracytoplasmic sperm injection techniques. Moreover, biomechanical principles of morphogenesis are appearing on the horizon of developmental biology, particularly in investigations aimed at understanding how and where physical forces are generated during epiboly in *Danio rerio* (Hernández-Vega *et al.*, 2017) or how mechanical forces are fundamental to the development and homeostasis of *Drosophila* in that they allow recognizing interactions of cell mechanics with cell growth and division during embryogenesis (Paci and Mao, 2021). These concepts ended up permeating the field of oncology, where the importance of cancer mechanoregulation was widely recognized (Chowdhury, Huang and Wang, 2022).

If we apply ideas from the field of physics (Campbell, 1999; Forgacs and Newman, 2005), which tend to be substantially underestimated in experimental biology, it can be argued that the cell surface area/volume ratio would be a fundamental element to explain the cell division process of the first embryo. With regard to this point, consolidation of synthetic biology has been crucial in establishing independent explanatory models for genes. For example, the use of giant vesicles with sizes comparable to those of cells revealed the propensity of vesicles to divide into spherical parts connected by small, narrow tubes (Božič and Svetina, 2007; Svetina, 2009). Other interesting models showed that changes in viscosity properties of synthetic membranes are required for asymmetric division, a characteristic observed in stem cells (Sato *et al.*, 2013) and in the third oblique cleavage of ctenophores. Studies on prokaryotic cells demonstrated that simple biophysical processes could have produced efficient conditions for cell proliferation during the evolutionary history of early cells, serving as biological models for studying this phenomenon (Mercier, Kawai and Errington, 2013). Excessive membrane synthesis in *Bacillus subtilis* seems to be sufficient to trigger cell division (Mercier, Kawai and Errington, 2013). Thus, an artificial increase in the cell surface area/volume ratio would have induced morphological changes and triggered proliferation. It is plausible to expect that all available properties of molecules and molecular complexes have been explored in the course of evolution (Svetina, 2009). It is therefore tempting to propose that today's complex life forms might be a more evolved version of syncytia formed by multiflagellate fusion of different life stages of unicellular holozoans, triggering the clonal proliferation process that we now call the embryo.

It is key to consider two aspects of the fusion of unicellular organisms: (i) they either merge to form a syncytium, without altering their life cycle, as occurs in sex cells of *D. discoideum* (McConachie and O∅Day, 1986), or (ii) they merge and divide similarly to the first animal embryo. Fusion and division of a syncytium entail additional complexity. Which nuclei will form the zygote, and how will the other nuclei participate in the process? This additional complexity seems like an insurmountable barrier, but protists themselves have already anticipated a partial solution to the problem in the form of simultaneous nuclear differentiation. In Protozoa, this process is found only in cases of nuclear dimorphism: there are generative nuclei capable of reproduction and "somatic" nuclei that are unable to reproduce or capable of reproducing only for a limited period, being destined to die thereafter (Grell, 1973). In *Rotaliella heterocaryotica*, the somatic nucleus (agamont) temporarily swells until becoming pycnotic and disintegrating in the cytoplasm (Grell, 1973). In *Rotaliella roscoffensis*, the agamont not only swells but also undergoes chromosomal condensation and even intranuclear spindle formation (Grell, 1973). So far, nuclear dimorphism has been shown to occur only in some groups of protozoans, foraminiferans, ciliates (Grell, 1973), and radiolarians. The primary nuclei of radiolarians (which are similar to the polyploid macronuclei of ciliates) and their polygenic nature are very controversial topics, even though it is clear that these structures are the only protozoan nuclei with capacity for multiple division (Grell, 1973). Radiolarian studies are substantially fragmentary because of the difficulty in collecting specimens from the greatest ocean depths. Another problematic factor is that these organisms are often infected by parasitic dinoflagellates, which multiply and simulate the multiple division capacity of the host (Grell, 1973).

An important feature of cell fusion is its fundamental role in protist sexuality, particularly in autogamy, gamontogamy, and gametogamy (Grell, 1973). In some groups of protists, gamontogamy is essential before the beginning of meiosis for the formation of sex cells (Grell, 1973). In cases of gametogamy, fusion is inherent to the sexual process. Some other cases, however, do not involve gamete fusion, but rather the formation of a fertilization tube for transmission of the gametic nucleus (Bhaud, Soyer-Gobillard and Salmon, 1988). In some cases of gametogamy, there is evidence of natural fusion of microgametes, with meiotic intermediates forming 3N or 4N polyploid cells in *Trypanosoma brucei* (Gibson, Garside and Bailey, 1992; Hope *et al.*, 1999; Gibson *et al.*, 2008; Peacock *et al.*, 2021), which served as an inspiration for my proposal of multiflagellate fusion with intermediate meiosis (Figure 1 and 2). The observed frequency of fusion events between gametes and meiotic intermediates would not conflict with a model involving meiosis (Gibson *et al.*, 2008; Peacock *et al.*, 2021) and its long evolutionary history. Perhaps the most important reflection is that trypanosomes, as members of the Excavata, carry an ancestral evolutionary history (Sogin, 1989; Sogin *et al.*, 1989; Dawson and Paredez, 2013) that diverged 1510ó1699 Ma (Parfrey *et al.*, 2011; Strassert *et al.*, 2021), about 1000 Myr before the beginning of oxygenation and the transition of oceans to a more modern chemical state.

Thus, the model of multiflagellate fusion proposed herein would involve a situation similar to nuclear dimorphism, but in the context of a haplontic cell at the beginning of meiosis (where the zygote would be the equivalent of a macrogamete) fused with several flagellated microgametes to generate a syncytial multinucleated primordial cell (Figure 1). Microgametes acted as stressors, momentarily interrupting meiosis before the release of the first polar body from this first syncytial multinucleate cell. In this manner, multiflagellate fusion permeated embryonic development during meiosis, triggering a series of structural changes that allowed embryogenesis to begin (Figure 2). Excess calcium in the Ediacaran ocean contributed to the consolidation of actin microfilaments, cell viscoelasticity, and fusion mechanisms. This particular moment of embryogenesis would be reproduced by germ cells of the first embryo, selecting meiosis as the specific type of cell division of this metazoan cell line.

**The restart of meiosis, co-option of the NFM, and consolidation of a syncytial multinucleated zygote**

Currently, there is a robust body of evidence indicating that the actin cytoskeleton assumes several essential functions during oocyte meiosis (Uraji, Scheffler and Schuh, 2018) and embryonic development (von Dassow, 1994; Guild *et al.*, 1997; Maniotis, Chen and Ingber, 1997), many of which rely on microtubules in mitotic cells (Reinsch and Gonczy, 1998). More specifically, the actin network is associated with the positioning of the nucleus within cells (Uraji, Scheffler and Schuh, 2018) and migration of the meiotic spindle to the cellular cortex, which is critical for asymmetric division during meiosis (Azoury *et al.*, 2008; Schuh and Ellenberg, 2008). Thus, the actin network participates in the formation of polar bodies (Binyam, Scheffler and Schuh, 2018) as well as in the precise segregation of chromosomes in mammalian oocytes (Mogessie and Schuh, 2017; Dunkley, Scheffler and Mogessie, 2022). It is also known that nuclear actin filaments are required for chromatin decondensation (Baarlink *et al.*, 2017; Moore and Vartiainen, 2017; Parisis *et al.*, 2017). Premature chromatin condensation is observed in models of cancer cell fusion (Williams, Scott and Beck, 1976; Atkin, 1979; Kovacs, 1985), which leads us to ponder on the notion that even these intricate details of prezygotic organization of the embryo were perfectly recruited and organized at that crucial moment in which multiflagellate fusion gave rise to the first animal.

As for events taking place after multiflagellate fusion, it should be noted that the nuclear membrane of spermatozoa (or flagellated microgametes in the current model) vesiculates and disintegrates (Longo and Kunkle, 1978; Poccia and Collas, 1997) (supporting the idea of polyploidy), allowing cytoplasmic proteins to come into contact with the genetic material of gametes (Garbers, Janette Tubb and Kopf, 1980; Porter and Vacquier, 1986; Perreault, Barbee and Slott, 1988). I speculate that this first syncytial cell, eminently sensory owing to its multiflagellate origin (with voltage-sensitive calcium channels), in a calcium-rich Ediacaran ocean (Sawaki *et al.*, 2014; Tostevin *et al.*, 2019), pre-equipped with a vast network of (nuclear and cytoplasmic) actin filaments (Soyer, 1981) connecting the plasma membrane and nucleus (Frieden and Gatenby, 2019), capable of transmitting calcium waves and ion signals through the actin cytoskeleton (Hunley and Marucho, 2022), was able to modify and regulate the genomic organization of the nucleus (Maniotis, Chen and Ingber, 1997).

Coherently, in metazoans, fertilization is a fusion phenomenon linked to calcium waves and genomic reprogramming (Xia and Xie, 2020) that impacts cell proliferation, adhesion, differentiation, and morphogenesis. Spontaneous cell reprogramming induced by cell fusion has been discussed and cited in this article. In tissue engineering models, chemical and mechanical stimuli exerted on mesenchymal stem cells promote increases in calcium channel activity, cytoskeleton reorganization, signaling of Ras/Raf/MEK/ERK cascade signaling, and simultaneous regulation of cell adhesion, proliferation, and differentiation (Liu *et al.*, 2013). Thus, I propose that this multiflagellate syncytial cell, via mechanotransduction, mechanosensing, and chromatin organization functions and apparently together with the nuclear lamina (as will be seen later in this section), managed to co-opt and induce the NFM and restart meiosis, release polar bodies, allow syngamy, and thus form the first syncytial multinucleated embryonic zygote. It is important to highlight that one of the nuclei of this multiflagellate cell had initiated meiosis; therefore, this sequence of events (from resuming meiosis to the zygote) is possible within a haplont cell impregnated with several flagellated microgametes.

A topological model of gene regulation, in which nuclear organization influences gene expression, has been proposed (Cremer and Cremer, 2001; Sexton *et al.*, 2007; Lieberman-Aiden *et al.*, 2009; Rao *et al.*, 2014; Gonzalez-Sandoval and Gasser, 2016), although the model does not interconnect the nucleus to the cytoplasmic context or considers the physical forces that the cytoskeleton transmits/exerts on the nucleus (Alam *et al.*, 2014; Li *et al.*, 2014; Kim and Wirtz, 2015; Arsenovic *et al.*, 2016; Kim *et al.*, 2017; Nozawa *et al.*, 2017). Protozoans have a prototype for the transmission of mechanical force and transduction of chemical signals into the nucleus, which is consistent with a microfilament-organizing center at the center of the cell (Mylnikov *et al.*, 2019) and with nuclear actin cytoskeleton (Jockusch, Brown and Rusch, 1971; Katsumaru and Fukui, 1982; Kumar *et al.*, 1984; Soyer-Gobillard, Ausseil and Géraud, 1996) or intranuclear actin microfibrils (Soyer, 1981). Therefore, protists can transmit mechanical force to the nucleus and alter its topological structure (Benken and Sabaneyeva, 2011) and chromatin condensation (Rubin, Goldstein and Ko, 1978). This supports the modern concept of actin being involved in chromatin remodeling, a process present in protists and that evolved in multicellular organisms (Jockusch *et al.*, 1974; Berdieva *et al.*, 2016). The concept is also consistent with my proposal for the NFM induced by multiflagellate fusion. Currently, it is recognized that mechanotransduction and mechanosensory systems directly modulate the dynamics of chromatin organization (Zheng *et al.*, 2009; Fedorchak, Kaminski and Lammerding, 2014; Cho, Irianto and Discher, 2017; Miroshnikova, Nava and Wickström, 2017; Uhler and Shivashankar, 2017b, 2017a; Wang *et al.*, 2017). I speculate that, in this manner, under the influence of physical phenomena, the first elements of the NFM were co-opted.

Modeling of nuclear organization and gene expression by mechanotransduction mechanisms is crucial for my hypothesis, as the neoplastic module is built sequentially and co-opts different developmental pathways while the embryo undergoes transformation during morphogenesis. Physical forces might have been fundamental in translating morphogenetic movements into a different nuclear topology capable of driving cell differentiation. Spatial separation of chromatin according to parental origin has been proven, which implicates that chromosomes would be in subcompartments within the nucleus in blastomeres of 2- and 4-cell mouse embryos (Mayer, Smith, *et al.*, 2000). The purpose of such spatial separation would be epigenetic reprogramming, which occurs differently in the two parental genomes (Rougier *et al.*, 1998; Mayer, Niveleau, *et al.*, 2000). Regarding nuclear compartmentalization, several studies revealed that chromosomes are organized within the nucleus in a tissue-specific manner (Lanctôt *et al.*, 2007; Bonev and Cavalli, 2016; Dekker and Mirny, 2016; Maharana *et al.*, 2016) and that there is a mechanical control of cell differentiation (Nelson, 2022).

Surprisingly, nuclear morphogenesis has been hypothesized, whereby the nucleus would have the ability to coordinate several lineage-specific transcriptional factors, spatially and temporally, facilitating gene regulation, ensuring cell differentiation, and promoting the emergence of robust morphogenesis (Tsai and Crocker, 2022). This hypothesis is completely compatible with my proposal of mechanotransduction imbued in NFM co-option. Thus, "nuclear morphogenesis" may be a microscopic counterpart to the physical process that shapes the animal body. In line with the promotion of spatial coordination by mechanosensory systems, an internal nuclear membrane protein (TMEM201) is capable of binding and phosphorylating SMAD2/3 and mediating nuclear translocation and transcriptional activation of TGF-β (Kong *et al.*, 2022). The presence of nuclear tyrosine kinase receptors (Chen, Hsu and Hung, 2020) and nuclear fibroblastic growth factor receptors (Suh *et al.*, 2022) in human cancer models point toward a broad mechanosensory mechanism in embryogenesis in the context proposed by my hypothesis.

From an evolutionary point of view and consistent with excess calcium in the Ediacaran ocean, which favors F-actin formation, F-actin has a relevant role in nuclear envelope breakdown (NEBD), also known as GVBD, of oocytes in meiosis I of *Patiria pectinifera* and *Patiria miniata* (starfish oocytes) (Mori *et al.*, 2014; Wesolowska *et al.*, 2020). Given the similar arrangement of F-actin in other animals (Jacobsohn, 1999; Burkel, von Dassow and Bement, 2007; DuBuc *et al.*, 2014), it is strongly suggested that NEBD is a widely used mechanism in animal evolution, but not the only one (Wesolowska *et al.*, 2020). It is also known that nuclear lamins provide nuclear stability, help connect the nucleus to the cytoskeleton, and may modulate chromatin organization and gene expression (Davidson and Lammerding, 2014). At a whole-cell level, lamins are involved in cytoskeletal organization, cell motility, and mechanotransduction (Vahabikashi *et al.*, 2022). Expression of different lamin isoforms has been associated with cell development, differentiation, and tissue-specific functions (Vahabikashi *et al.*, 2022). With regard to cell differentiation, nuclear lamin expression and function are associated with large-scale organization of chromatin domains (Bitman-Lotan and Orian, 2018, 2021; Alcorta-Sevillano *et al.*, 2020). All of these functions are very similar to those of actin filaments within the nucleus. There is evidence that nuclear lamins and actin filaments interact physically (Simon, Zastrow and Wilson, 2010). Such convergence and overlapping of functions refer us to the beginning of recruitment of the neoplastic module. It most likely would have occurred at the time of nuclear envelope rupture, at the beginning of meiosis I; apparently, GVBD was also co-opted to Neoplasia and physical mechanotransduction systems.

Lamins translate mechanical stimuli into biochemical signals (mechanosensors) (Irianto *et al.*, 2016; Cho, Irianto and Discher, 2017; Xia *et al.*, 2018; Xie, Walker and Irianto, 2020) involving NF-κB signaling pathways (Lammerding *et al.*, 2004, 2005). Lamin A/C associates with at least 54 transcriptional regulators, including c-Fos, ERK1/2, SREBP1, and pRb (Simon and Wilson, 2013). In line with the proposed hypothesis, an increasing number of reports have implicated nuclear lamins in human cancers (de las Heras, Batrakou and Schirmer, 2013). For example, rupture of the nuclear envelope occurs particularly in micronuclei and is preceded by local rupture of the lamina (Hatch *et al.*, 2013). In the case of colorectal cancer, increased and decreased levels of lamin A/C have been shown to be correlated with increased aggressiveness, and reduced levels are associated with tumor recurrence in patients at advanced cancer stages (Willis *et al.*, 2008; Belt *et al.*, 2011). According to some authors, changes in lamin expression may affect cancer progression through a variety of mechanisms, including altered proliferation, signaling, and cell migration (Kong *et al.*, 2012; de las Heras, Batrakou and Schirmer, 2013; Davidson and Lammerding, 2014; Denais *et al.*, 2016).

The viscoelastic physical properties of the cortex, at resumption of meiosis I, help elucidate the close connection between the actin cytoskeleton and nuclear dynamics. Furthermore, it suggests that evolution research should be focused on how the same proteins are used differently in different physical contexts. For example, anisotropic contraction of the cytoplasmic actin mesh in starfish oocytes is relevant in spindle migration during meiosis I (Mori *et al.*, 2011). In mammals, instead of contraction, the forces at play for spindle migration stem from the displacement of spindles via an elastic mesh of actin filaments (Schuh and Ellenberg, 2008). Two distinct physical phenomena occur with the same cellular structure. Detailed studies on mice showed that, after resumption of meiosis I, the actin cortex becomes thicker and an inner, less dense actin layer called the subcortex is formed. This cortical thickening is dependent on the activity of the Arp2/3 complex, a nucleator of branched actin networks (Azoury *et al.*, 2008; Chaigne *et al.*, 2013). The mechanics of the cortex depend on cortical exclusion of myosin II, which decreases cortical tension and transduces tensile forces from spindle poles to the cortex, promoting migration (Chaigne *et al.*, 2013). Moreover, balance of this tension force has been shown to be essential for such migration (Chaigne *et al.*, 2015). This behavior is completely opposite to that of mitotic cells, for which increased cortical tension is crucial for controlling the orientation of the mitotic spindle (Fink *et al.*, 2011). Finally, cortical thickening and its textural properties are fundamental. By producing artificial stiffness using concanavalin A, it was possible to inhibit 80% of cortical spindle migration; as a result, spindles remained at the center of the oocyte. This means that a soft cortex is required for spindle migration in meiosis I. For an extensive review of the role of the actin cytoskeleton in oocytes and zygotes see Uraji, Scheffler and Schuh, 2018 e Dunkley, Scheffler and Mogessie, 2022.

It can be predicted that, after formation of the syncytial multinucleate cell, resumption of meiosis and migration of the spindle occurs, and, according to the clues given by nuclear dimorphism of protists (Grell, 1973), many nuclei of flagellated microgametes become pyknotic and only one of them will participate in syngamy. It can also be predicted that excess $Ca^{2+}$ in the Ediacaran ocean favored the formation and contraction of actin microfilaments (Taylor *et al.*, 1973) in detriment to the formation of microtubules (Schliwa *et al.*, 1981). This allows us to affirm that, under the excess calcium conditions of the Ediacaran ocean, the role of molecules such as calmodulin must have been essential (Schliwa *et al.*, 1981) for cells to utilize microtubule cytoskeleton networks.

In this sense, it is important to show that the actin filament network is not universal in evolution, and other cytoskeleton networks are used for similar functions in different organisms. In *Drosophila melanogaster*, for instance, the actin network disappears before meiotic maturation (Dahlgaard *et al.*, 2007), and microtubules seem to play a preponderant role in the process. Initially, the asymmetric positioning of the meiotic spindle in *X. laevis* appeared not to depend on F-actin (Gard, Cha and Roeder, 1995). There is strong evidence of an interaction between the microtubule cytoskeleton and some components of the actin cytoskeleton (Waterman-Storer *et al.*, 2000) and of a collaboration between the two to enable spindle positioning and orientation, which is important for asymmetric division in *Xenopus* meiosis (Weber *et al.*, 2004). Interaction between the actin cytoskeleton and microtubules has also been described in mice (Azoury *et al.*, 2008), and the role of centrosomes as actin organizing centers has been recognized (Farina *et al.*, 2016).

Finally, according to my hypothesis, the multinucleated syncytial cell had two forces to divide, the neoplastic and the physical force, determined by the surface volume relationship. In addition, the syncytial zygote had many pyknotic nuclei (DNA) that did not participate in syngamy, disintegrating in the cytoplasm of this primordial cell.

**Division of the zygote and impact on polyploidy**

Considering the life cycles of unicellular holozoans, it is deemed necessary that three forms of nuclear division be controlled at the beginning of embryogenesis, so that there is no interference with zygote division: (i) mitosis, which was privileged in the course of embryonic formation and multicellularity (in the context of this article, it is called proliferation), (ii) meiosis, which was conducted for germ line and endomitosis control, and (iii) endomitosis (a type of meiosis), which produces polyploidy in animals and is very important in the formation of coenocytes in unicellular holozoans (syncytium-like structure according to some authors) (Ros-Rocher *et al.*, 2021). Because it is at the base of animal phylogeny, endomitosis has become a transcendental event also for animal embryogenesis (Anatskaya and Vinogradov, 2022). On the other hand, endoreplication control by physical mechanotransduction mechanisms is well known to occur in embryogenesis and animal physiology (Luff and Papoutsakis, 2016; S. Wang *et al.*, 2018; Uroz *et al.*, 2019). That is, at the beginning of animal embryogenesis, mechanotransduction mechanisms controlled these three nuclear division processes, which were co-opted to the NFM. The development of this part of my hypothesis was inspired by the wonderful work of Caroline Uhler and by imagining how the three-dimensional organization of the embryo could contribute to understanding the organization of the cell nucleus and emergence of cancer (Uhler and Shivashankar, 2017b, 2018; Wang *et al.*, 2017; Belyaeva *et al.*, 2022).

This type of syncytial multinucleate organization, which integrated and recruited the NFM, including endomitosis control, impregnated embryogenesis in all groups of animals. This means that polyploidy has become an embryological, cancerous, and, thus, evolutionary phenomenon. Examples of mononuclear or multinucleated giant cells are commonly observed in human embryos (Pickering *et al.*, 1995; Kligman *et al.*, 1996) and, because they are linked to aneuploidy and other chromosomal aberrations, have been associated with non-viability of embryos (Vanneste *et al.*, 2009; Vera-Rodriguez *et al.*, 2015) or chaotic developmental events (Ledbetter, 2009; Liu, 2018). In *D. melanogaster*, polyploid cells allow regenerating functional stem cells in the intestine (Lucchetta and Ohlstein, 2017), and, in mammals, polyploidy is a very common feature of the liver (Gentric and Desdouets, 2014), related to the survival of heart and liver tissues (Anatskaya and Vinogradov, 2007). Surprisingly and in agreement with our functional module, mice and human transcriptome analyses demonstrated that somatic cell polyploidy can regulate the organism's developmental gene modules, increasing their activity (Anatskaya, 2018). In other words, polyploidy is incorporated into morphogenetic and embryological processes under physiological conditions (Anatskaya, 2018). Consistent with my proposal of an NFM recruited by a syncytial multinucleate organization (multiflagellate fusion), somatic polyploidy is associated with upregulation of genes that interact with Myc and epithelial–mesenchymal transition (Vazquez-Martin *et al.*, 2016), two important components of my theoretical framework. In my proposal, it is the nuclei impregnated during multiflagellate fusion that induced the co-option of NFM components, thereby producing a polyploid syncytial multinucleate zygote. Among the co-opted and highly regulated components are those responsible for endomitosis division, used by our unicellular holozoan predecessors to produce polyploid multinucleated cells during normal embryogenesis. The multinucleated zygote and multinucleated cells of embryogenesis are an incredible coincidence for the beginning of animal phylogeny.

Opposite to chaos, polyploidy is related to a normal physiological condition (Vitale *et al.*, 2011; Gjelsvik, Besen-McNally and Losick, 2019; Donne *et al.*, 2020; Anatskaya and Vinogradov, 2022) as a biological force that communicates with biodiversity, biocomplexity (Fox *et al.*, 2020b), and, particularly, evolutionary significance (Van De Peer, Mizrachi and Marchal, 2017; Van de Peer *et al.*, 2021). In line with the hypothesis of cancer and embryo as two faces of the neoplastic process, numerous scientific studies have shown the spontaneous presence of polyploid multinucleated cells in solid tumors (Levan and Hauschka, 1953; Zack *et al.*, 2013; Fei *et al.*, 2015; Mu *et al.*, 2017). In paclitaxel-induced polyploid models of ovarian cancer, cells transformed after division into phenotypes resembling human blastomeres, expressing embryonic stem cell markers (Oct4, stage-specific embryonic antigen 1, Sox2, and NANOG) and producing somatic and germ lines (Niu, Mercado-Uribe and Liu, 2017), consistent with *in vitro* activation of our NFM. Polyploid multinucleated cells were able to induce teratomas and form tissues equivalent to the three embryonic germ layers (Niu, Mercado-Uribe and Liu, 2017). An embryonic theory of cancer origin based on polyploid multinucleated cells has been formulated (Liu, 2020), but it does not contemplate one of the main aspects of the modern cancer theory, namely the evolutionary perspective, given that cancer affects all animal groups (Sparks, 1972; Kaiser, 1989; Rothschild, Witzke and Hershkovitz, 1999; Robert, 2010; Tascedda and Ottaviani, 2014; Aktipis *et al.*, 2015; Sinkovics, 2015; Albuquerque *et al.*, 2018). A theory with an evolutionary point of view was developed by Vladimir Niculescu (Niculescu, 2021). It highlights the very inspiring similarities between polyploid multinucleated cells of protists (Niculescu, 2018a, 2018b, 2018c, 2020) and polyploid multinucleated cells produced spontaneously within a tumor. The theory is formulated within a framework of atavistic theories but does not address an embryonic perspective of cancer. On the other hand, my hypothesis brings together embryonic, atavistic, and evolutionary views and contemplates the role of physics in the formation of the first embryo and cancer.

Finally, there are two important questions to answer. Who regulates endomitosis? What is its evolutionary significance? Modified meiosis (without karyogamy) evolved in protists to mitigate the instability of polyploidy (Cleveland, 1947). As reported by Lemuel Cleveland, polyploidy in *Barbulanympha* is invariably reduced by meiosis without karyogamy (Cleveland, 1947). Other groups such as ichthyosporeans perform the opposite: they produce polyploidy by coenocytic division (nuclear division without division of the cytoplasm) (Ros-Rocher *et al.*, 2021). Surprisingly, injection of Mos mRNA at low concentrations into two-cell embryos of *Clytia hemisphaerica* formed polyploid cells with multiple nuclei or spindles within a common cytoplasm. This is a demonstration of Mos-dependent endomitosis in Cnidaria (Amiel *et al.*, 2009). Blockade of Mos in oocytes produces parthenogenesis in *C. hemisphaerica* and *Asterina pectinifera* (starfish) (Tachibana *et al.*, 2000; Amiel *et al.*, 2009), showing that Mos is also involved in targeting meiosis II and allowing ploidy reduction (Furuno *et al.*, 1994; Tachibana *et al.*, 2000; Amiel *et al.*, 2009). Polyploid multinucleated cells arise spontaneously in solid tumors (Levan and Hauschka, 1953; Zack *et al.*, 2013; Fei *et al.*, 2015; Mu *et al.*, 2017), and Mos could help return to a mitotic cycle of endopolyploid tumor cells through somatic reduction (Erenpreisa, Kalejs and Cragg, 2005). Studies on chemical induction of polyploidy in ovarian cancer produced cell clusters similar to blastomeres that segregate into germ and somatic lines (Niu, Mercado-Uribe and Liu, 2017). Cancers, as we know, recapitulate the germline (soma-to-germline transition) (J. Wang *et al.*, 2011; Feichtinger, Larcombe and McFarlane, 2014). Intuitively, if the control of endomitosis and meiosis is made by the same regulatory protein, we would expect cancer polyploidy to recapitulate the germline. In this sense, some authors suggest that somatic pairing, endomitosis, meiosis alterations, and chromosomal aberrations may all be correlated (Beçak, Beçak and Pereira, 2003). In the conceptual framework of my hypothesis, Mos controls the two types of meiosis, in the context of the NFM.

**Division in embryogenesis using the NFM**

With regard to the first cell divisions of the embryo, the viscoelasticity properties of a multicellular structure of 4, 8, or 16 cells are determined not only by distribution of adhesion molecules on the cell surface, which sustain the response of an external tension, but also by the chemical nature that allows this interaction between plasma membranes. The major classes of cell adhesion molecules, such as cadherins, are calcium-dependent (Takeichi, 1995; Gumbiner, 1996), as are various other molecules (Forgacs and Newman, 2005). In the context of the beginning of multicellularity, it would not be useful to have a protein repertoire with cadherins in a calcium-deficient environment. Furthermore, the property of viscoelasticity in response to an external force has been shown to be relevant in all mechanotransduction phenomena, some of which are described here. Biophysical phenomena are behind key aspects of embryo formation, such as maintenance of cell adhesion, which allows cells to stay together, and viscoelasticity, which allows a response to the mechanical microenvironment, reflecting on the orientation of the mitotic spindle and cell proliferation.

Here, I reflect on the asymmetric division of meiosis I and II, which, in my view, affects the somatic cells of the embryo. Intracytoplasmic displacement of the meiotic spindle in the direction of the cellular cortex seems to be decisive in the formation of unequal cells that form the polar body (Verlhac *et al.*, 2000; Azoury *et al.*, 2008; Chaigne *et al.*, 2013). The more central the nucleus is in the cell, the more egalitarian the division of the cytoplasm. It is well known that, in ctenophores, separation between the soma and germ line is fluid (Edgar, Mitchell and Martindale, 2021). In other words, multipotent cell populations are established, whose destinations include somatic cells and the germline (Juliano, Swartz and Wessel, 2010; Fierro-Constaín *et al.*, 2017). Actual separation of the germline occurs at some time during embryogenesis (Seervai and Wessel, 2013). Therefore, I speculate that multipotent embryonic cells can control meiosis, mitosis, and endomitosis, because they are contained within the neoplastic module. I further speculate that embryos would use meiotic mechanisms to produce the asymmetry of the third oblique division, triggering the formation of the circular halo of ctenophores, responsible for epiboly movements, which is considered to be the first great revolution of animals.

Recent studies described how fibronectin geometric patterns induce polarization of subcortical dynamic actin structures that correlate with mitotic spindle movements (Fink *et al.*, 2011). It is impossible to imagine that mechanical phenomena of adhesion and tension do not also involve the extracellular matrix. Thus, the extracellular matrix appears to control the location and dynamics of actin in contact with the membrane, influencing the segregation of cortical components at the interphase (Théry *et al.*, 2005). This segregation is subsequently maintained in the cortex of mitotic cells and used for spindle orientation during cell division (Théry *et al.*, 2005). Coherently, recent data support that the orientation of cell division in mammals (Lehtonen, 1980) and sea urchins (Masui and Kominami, 2001) appears to be defined by cell adhesion and tensile forces developed during the interphase. Therefore, nuclear and mitotic spindle positioning are dynamic events that depend on the forces and torques applied by the cytoskeleton and molecular motors (Rappaport, 1996; Minc and Piel, 2012).

Finally, the shape of a cell in a multicellular context would be the result of adhesive and tension patterns. These patterns control the spatial distribution of cortical signals and, surprisingly, aid spindle orientation and daughter cell positioning during proliferation. Therefore, cell division seems to be a continuous transformation that ensures the maintenance of tissue mechanical integrity (Théry and Bornens, 2006). Cells divide according to clues provided by their mechanical microenvironment, always aligned with an external force field (Fink *et al.*, 2011).

In my view, the Ediacaran ocean was decisive for syncytium formation, transformation of physical properties in the cell, and proliferation of cells that were kept together and organized by the mechanical environment of the first embryo. It may also be predicted that one of the first important events in animal evolution from ctenophores was the incorporation and association of cytasters of fused cells into nuclear membranes, representing evolutionary records of multiflagellate fusion.

**The first embryo was a hermaphrodite ctenophore without tentacles: An embryological view supporting ctenophores as the most basal animal lineage**

Although the traditional view is that sponges are the first branching lineage of the animal tree (Linden and King, 2021), the evolutionary importance of ctenophores has recently been highlighted. Several studies have suggested placing Ctenophora as the most basal animal lineage (Ryan *et al.*, 2013; Moroz *et al.*, 2014; Arcila *et al.*, 2017; Shen, Hittinger and Rokas, 2017; Whelan *et al.*, 2017). Most research on animal phylogeny requires a complex pattern of convergent evolution of animal tissues and systems, and the placement of ctenophores as a sister group to all other metazoans raises the possibility that the muscles and nervous system of ctenophores and other complex animals evolved independently (Moroz *et al.*, 2014).

Extant ctenophores possess a diploblastic body plan organization and some traits shared with bilaterians, including a functional intestine with two anal pores and muscles (Martindale and Henry, 1999). The nervous system of extant ctenophores consists mainly of mesogleal and ectodermal nerve networks (Jager *et al.*, 2011; Norekian and Moroz, 2019a, 2019b), and this pattern can be complemented with functional specializations related to feeding, such as prominent nerves associated with tentacle pairs (Jager *et al.*, 2011) or sealing of the mouth in beroids (Tamm and Tamm, 1991, 1993). The genus *Euplokamis* is unique among extant representatives in that it has "giant axons" (Mackie, Mills and Singla, 1992; Norekian and Moroz, 2020). On the other hand, morphological phylogenetic analyses conducted in the last century showed that the common ancestors of ctenophores would lack tentacles (Harbison, 1985). Recent studies of fossil records point in the same direction, suggesting that the first ctenophore was very similar to individuals of the order Beroida (atentaculate), underscoring an astounding level of nervous system sophistication that surpasses that of *Euplokamis* (Parry *et al.*, 2021).

From a sexual point of view, ctenophores are almost all hermaphrodites (individuals capable of producing male and female gametes during their reproductive cycle) (Pianka, 1974). Protandry (dichogamous hermaphroditism), in which male sex organs mature before female ones, occurs in Platyctenida (Pianka, 1974). Gonochorism or dioecy is only found in two species of the tentaculate genus *Ocyropsis* (Harbison and Miller, 1986). This argument explains, from an embryological point of view, why ctenophores, like beroids, could have been the first animals. There would be no possibility of producing a sexually differentiated animal at the beginning of animal phylogeny. That is, there would be no way to form a female and male embryo originating from a first fusion event ("fertilization") based on the fusion of unicellular holozoan cells. Although fungi with unisexual reproduction capacity exist (Roach *et al.*, 2014; Heitman, 2015), they do not fulfill the main requirements for the emergence of metazoans, namely anisogamy and a life cycle with flagellated amoeboid cells.

While it is difficult to rationalize the hermaphroditism of early multicellular forms, the standard impulse, in Stuart Newman's words, to conceptualize the emergence of biological novelties as the result of a competitive game originates from an assumption deeply rooted in neo-Darwinism (Newman, 2016). Thus, most authors manage to see the "advantages" of hermaphrodite sexuality (Ghiselin, 1969; Bachtrog *et al.*, 2014) at the expense of thinking that hermaphroditism is an inevitable consequence of the formation of the first embryo. Here, I refer specifically to the genetic (transmission advantage but with inbreeding depression) (Fisher, 1941; Lande and Schemske, 1985) and ecological (guarantee of reproduction and variable availability of partners) aspects of self-fertilization (Jarne and Auld, 2006; Munday, Buston and Warner, 2006). When it is difficult to find a reproductive partner, it is "advantageous" for individuals to be simultaneously male and female. This condition would allow them to mate with any other individual they encounter or even with themselves. The "need" for a guarantee of reproduction would be one of the reasons why plants, which cannot actively search for a partner, are hermaphrodites (Futuyma and Kirkpatrick, 2017).

It was this neo-Darwinian way of thinking that ingrained the notion that to consider ancestral metazoans as hermaphrodites would constitute erroneous scientific reasoning (Ghiselin, 1969). What is most surprising about this view is that the genetic models themselves predict that convex fitness conditions would be an "optimal resource allocation for a hermaphrodite" (Charnov, Bull and Maynard Smith, 1976). Although these authors did not consider the factors of "sperm storage" or "sperm competition" in their model (Charnov, 1979), let me raise the following question: in considering the first and only animal, what would be the degree of competition if it exists alone? On the other hand, in this initial condition of few animals in the Ediacaran ocean, as predicted by my hypothesis, all conditions of convex fitness are met; a low density of animals and a low overlap of resources in the production of male and female gametes would be expected (Charnov, Bull and Maynard Smith, 1976; Heath, 1979). Therefore, according to genetic models, a hermaphroditic ancestor would be expected at the base of metazoans. Furthermore, there is no logical need to think that multicellular forms would have to be competitively superior to unicellular forms for the former to endure (Newman, 2016), and there is no need to appeal to competition to explain survival in new ways, even if appearing abruptly (Newman, 2016). This last point is crucial to my hypothesis and radically opposes the theory of natural selection, which predicts that any major transition or large-scale change would occur gradually and based on relative fitness. Early ctenophores formed abruptly and had an even more sophisticated nervous system than that of extant ctenophores, as supported by fossil records (Parry *et al.*, 2021).

Finally, hermaphroditism is a very common characteristic in multicellular organisms, especially among plants and animals (Futuyma and Kirkpatrick, 2017), and separate sexes may evolve from hermaphrodites (Ashman, 2002). A partial solution is the spatial separation of male and female gonads in the same individual, as occurs in monoecious plants with separate male and female flowers or in most hermaphroditic animals (Futuyma and Kirkpatrick, 2017). Alternatively, male and female functions may be temporally separated within an individual, as found in many plants and some animals (Munday, Buston and Warner, 2006), such as ctenophores of the order Platyctenida (Pianka, 1974; Glynn *et al.*, 2019). Male and female reproductive organs may also be segregated into different individuals, as in ctenophores of the order Lobata, genus *Ocyropsis* (Harbison and Miller, 1986). The fact that this group of oceanic ctenophores evolved dioecy directly contradicts the claim that there is a selective advantage of hermaphroditism in environments where the likelihood of finding a mate is reduced (Harbison and Miller, 1986). The factors, including epigenetic ones, that are actually involved in the separation of sexes can be found, in my view, in the relationship of the embryo with its environment.

**Ediacaran ocean**

*Calcium*

Since Darwin's theory, biologists and scholars of Earth history have been fascinated by the abrupt appearance and rapid diversification of primitive animal life. However, the lack of data and consensus on the Ediacaran Period has limited discussion on metazoan evolution (Cloud and Glaessner, 1982). Today, it is thought that metazoan evolution was facilitated by a series of dynamic and global changes in redox conditions and nutrient supply, which sustained multiple radiation phases in metazoans, one of them in the Ediacaran Period (Wood *et al.*, 2019). The important role of environmental conditions in the emergence of embryogenesis and evolution of living forms is also becoming increasingly clear (Moczek, 2015). In this context, $Ca^{2+}$ arose as a promoter of the main stages of metazoan evolution. Thus, it has been proposed that changes in $Ca^{2+}$ concentration in the Ediacaran ocean were a crucial driving force behind major innovations in evolution, including multicellularity and the origin of metazoans (Brennan, Lowenstein and Horita, 2004; Kazmierczak, Kempe and Kremer, 2013)

The content of cytosolic $Ca^{2+}$ in all non-excitable cells (at rest) is kept constant at about $10^{7.5}$ M, being therefore lower by several orders of magnitude than that in extracellular fluids, such as blood or seawater ($10^{-2}$ to $10^{-4}$ M) (Kretsinger, 1983). Concentrations above $10^{-7}$ M are considered deleterious to cellular function (Kazmierczak, Kempe and Kremer, 2013), showing that today's seawater, which contains 400 mg/L, is lethally toxic to cytosol functioning (Goldberg, 1963; Cotruvo, 2005). Furthermore, it is suggested that the intracellular medium is older than its external medium, as currently constituted. The relative proportions of inorganic elements in the cell are of older origin than the relative proportions of the same elements that prevail in the external environment or in today's ocean waters (Macallum, 1926). In this way, life originated in an environment with low $Ca^{2+}$ levels, similar to that of the cytosolic level, and, during evolution, adjusted to increasing levels of $Ca^{2+}$ and developed elaborate devices to enable survival (Macallum, 1926; Kazmierczak, Kempe and Kremer, 2013). This implies that the chemistry of the Ediacaran ocean, with regard to calcium, was completely different from that of modern oceans. High $Ca^{2+}$ concentrations must have imposed formidable stress but also expanded the possibilities for diversification and innovation of the evolving biota.

It is postulated that the ancient ocean had high alkalinity and low concentrations of $Ca^{2+}$ and $Mg^{2+}$, similar to modern soda lakes, commonly associated with volcanic regions (Kempe and Degens, 1985). It is assumed that the first soda ocean gradually evolved into a sodium chloride-rich ocean through "titration," emanating volatile acids such as HCl, thereby reducing the pH and finally reaching a titration point of $Ca^{2+}$, allowing its relatively rapid accumulation at the end of the Precambrian (Kempe and Degens, 1985; Kazmierczak, Kempe and Kremer, 2013). In a time period of 3 billion years, the primitive concentration of calcium in the ocean increased by 1,000 to about 100,000 times (Kempe and Degens, 1985; Kazmierczak, Kempe and Kremer, 2013). A strong indicator of an early alkaline ocean is the predominance of alkaloid cyanobacteria (Cyanophycea) in Precambrian biota (Kempe and Degens, 1985). The later appearance of metazoans and biological calcification can be interpreted as the result of a gradual increase in $Ca^{2+}$ during the final decay of the soda ocean near the Ediacaran Period (Cloud and Glaessner, 1982; Porter, 2007; Wood, 2011; Pruss *et al.*, 2018). The increase in stress promoted by $Ca^{2+}$ forced living systems to store or excrete excess $Ca^{2+}$, which would otherwise poison vital cellular processes (Elbrink and Bihler, 1975). Contrary to general belief, an extensive toolkit of protection and signaling by $Ca^{2+}$ was found in the unicellular choanoflagellate *M. brevicollis* (Cai, 2008).

*Oxygen and the Warburg effect*

There is still considerable debate as to whether or not oxygenation was the main factor in early metazoan evolution (Towe, 1970; Wood *et al.*, 2019). Although all existing animals need oxygen, their demands are not always the same. The most tolerant organisms of extreme oxygen depletion on the seafloor include calcareous foraminiferans, nematodes, and annelids (Levin, 2021). A recently developed biogeochemical model revealed that atmospheric and oceanic oxygen levels in the Ediacaran Period were dissociated, indicating that the early increase in atmospheric oxygen levels in the Ediacaran might not have promoted generalized oxygenation of the deep ocean (Shi *et al.*, 2022), which is also consistent with geochemical records (Lenton *et al.*, 2014). Experimental results with basal groups of animal evolution demonstrated that the sea sponge *Tethya wilhelma* can withstand oxygen levels as low as 0.25% to 4% of current atmospheric levels (Mills *et al.*, 2014). Ctenophores also live in deep ocean waters (Harbison, Madin and Swanberg, 1978; Haddock *et al.*, 2017) and some have been found at 7,200 m depth (Lindsay and Miyake, 2007), where there are extremely low oxygen levels. Genomic analyses of sponges and ctenophores showed that both lack important components of the hypoxia-inducible factor (HIF) pathways that maintain cellular oxygen homeostasis (Mills *et al.*, 2018). Therefore, ctenophores might have performed aerobic metabolism under very low concentrations of oxygen.

Surprisingly, studies on deep-sea creatures, such as the copepod *Tigriopus californicus*, showed that they can withstand prolonged exposure to extreme oxygen deprivation, despite having secondarily lost the main members of the HIF pathway, being similar to tumor systems (Graham and Barreto, 2019). Allie Graham and Felipe Barreto predicted a shift from oxidative phosphorylation to glycolysis under hypoxic conditions. Consistent with this prediction, the authors observed a significant increase in pyruvate dehydrogenase kinase (PDK) mRNA levels (Graham and Barreto, 2019). This is the Warburg effect (Warburg, 1956) occurring in deep ocean waters. Despite a large volume of papers researching this effect in cancer, the function of the Warburg effect has remained unclear (Liberti and Locasale, 2016). Another expected prediction for the Warburg effect is that elevated glucose metabolism decreases the pH within the cell due to lactate formation (Estrella *et al.*, 2013), thereby requiring greater activity of $H^+$ extrusion mechanisms to prevent cell death (DeCoursey, 2015). Considering the animal formed in the Ediacaran ocean under hypoxic conditions (Shi *et al.*, 2022) in an alkaline sea (Kempe and Degens, 1985), it is expected that it would show $H^+$ efflux. Voltage-gated proton channels (Hv1) with pH-dependent regulation occur in several organisms, from protists to humans (DeCoursey, 2015), and are responsible for bioluminescence in dinoflagellates (Rodriguez *et al.*, 2017; Kigundu, Cooper and Smith, 2018). These channels might have helped calcify (Taylor *et al.*, 2011; Kottmeier *et al.*, 2022) and acidify the Ediacaran ocean (external environment). Acidification of the microenvironment in tumor models has always been considered an intriguing aspect of the disease (Schlappack, Zimmermann and Hill, 1991; Rofstad *et al.*, 2006; Moellering *et al.*, 2008). The potential benefits of acidosis to cancer cells have been discussed. Some hypotheses of acid-mediated invasion suggest that $H^+$ ions secreted by cancer cells diffuse into the surrounding environment and alter the stroma interface of the tumor, allowing for greater invasiveness (Gatenby and Gawlinski, 1996; Estrella *et al.*, 2013). It is possible that the Warburg effect and acidification are a remnant of our evolutionary past linked to Neoplasia. Thus, in the first ctenophores, the Warburg effect might have promoted high proliferation and cell growth in a hypoxic and highly alkaline environment.

In line with these observations, the Warburg effect seems to be co-opted to the NFM. Disruption of Pdk1 expression decreased lactate production in cancer cells, suggesting elimination of the Warburg effect (Kim *et al.*, 2006; Papandreou *et al.*, 2006). Hypoxia-inducible Pdk1 is critical for attenuation of reactive oxygen species (ROS) production (Kim *et al.*, 2006); however, such mechanism does not make much sense in an animal arising in an environment with little or no oxygen and without HIF-1 (Mills *et al.*, 2018). Possibly, NADPH oxidase (Nox2)/ROS pathways and HIF-1/Pdk1 metabolic switch (Kim *et al.*, 2006) were co-opted later in animal evolution, during the oxygenation phase of oceans (Neoproterozoic oxygenation event) (Shi *et al.*, 2022) as a response to $Ca^{2+}$ fluctuations in the environment (sporadic anoxic events) (D. Wang *et al.*, 2018). Of the highly expressed hypoxia-related genes, Pdk1 is a potential target of Myc, as identified by chromatin immunoprecipitation experiments (Li *et al.*, 2003). Myc activates glycolytic genes, including lactate dehydrogenase, and is therefore thought to play a role in the Warburg effect (Kim *et al.*, 2004). Myc did not activate Pdk1 in the P493-6 human Burkitt's lymphoma cell line (Kim *et al.*, 2006). On the other hand, homologs of the Hv1 channel were characterized in ctenophores, indicating a very ancient origin in evolution (Moroz and Kohn, 2016). Regarding Neoplasia, recent studies suggest that Hv1 may contribute to the malignancy of various tumors (Y. Wang *et al.*, 2011; Wang *et al.*, 2012, 2013; Morgan *et al.*, 2015). Consistent with my hypothesis, Hv1 knockdown reduced cell proliferation, cell migration, and MMP release from the extracellular matrix in a breast cancer cell line (Morgan *et al.*, 2015). Such findings suggest the participation of Hv1 and a functional module in early evolution. Later in animal evolution, the Nox2/ROS system was recruited to the NFM in an environment with a considerable increase in oxygenation.

This functional module is most evident in B lymphocytes, demonstrating their incredible conservation and suggesting that co-option and redirection of pre-existing systems are the main source of evolutionary innovation (Litman, Rast and Fugmann, 2010). B lymphocytes are crucial in immune system responses in higher vertebrates, which use cellular strategies of invertebrates (Akira, Takeda and Kaisho, 2001; Sun and Lanier, 2009; Vinkler and Albrecht, 2011) that live in deep waters. The activity of B lymphocytes is regulated by signaling pathways that are fundamental for the commitment of these cells to proliferation, migration, and differentiation into plasma cells (Paus *et al.*, 2006). B lymphocyte signaling activates voltage-gated proton channels that facilitate ROS production (DeCoursey, Morgan and Cherny, 2003). Blockade of the Hv1 channel in B cells reduced ROS production and afforded a weaker response to antibodies *in vivo* (Capasso *et al.*, 2010). The prevailing view is that proton channels are needed to maintain pH in a range that allows NADPH oxidase activity. Hv1 is expressed in B cells of patients with chronic lymphocytic leukemia (Boyd, Dyer and Cain, 2010). A shorter Hv1 isoform showed less internalization, higher proton current, and greater proliferation and migration of B cells (Hondares *et al.*, 2014). These properties of the short isoform are consistent with the idea that excessive Hv1 activity promotes malignancy. We can see our evolutionary origin by looking at cancer. Excessive Hv1 activity in an alkaline environment is coherent with high cell proliferation and growth (neoplastic process). Inhibition of Hv1 induced apoptosis of metastatic glioma cells (Wang, Zhang and Li, 2013). A similar Nox2/ROS/Hv1 system is observed in granulocytes (Ramsey *et al.*, 2009; Zhu, Mose and Zimmermann, 2013) and myeloid-derived suppressor cells (Alvear-Arias *et al.*, 2022; Cozzolino, 2022), which proliferate in tumor microenvironments.

At present, it is very difficult to discuss apoptosis in ctenophores because Bcl-2 and Yorkie (an apoptosis modulator of Hippo pathways) (Huang *et al.*, 2005) were not found in genomic analyses (Ryan *et al.*, 2013; Moroz *et al.*, 2014; Coste *et al.*, 2016; Suraweera *et al.*, 2022). Therefore, I preferred not to comment yet on their presence in the NFM. Caspase 7 was found in *Mnemiopsis leidyi* and *Pleurobrachia bachei*, indicating that apoptosis can be initiated via a different pathway.

**Concluding remarks and perspectives**

One of the main aspects of this final reflection is that cancer reveals much of our evolutionary history. The Warburg effect takes on a different meaning in the context of the external conditions of the Ediacaran ocean. The Warburg effect and acidification seem to be a remnant of our evolutionary past, linked to Neoplasia, which allowed intense proliferation in a hypoxic and alkaline environment. Coincidence or not, 95% of ctenophores produce bioluminescence below 500 m depth (Martini and Haddock, 2017), possibly related to the low oxygenation of deep waters and acidification by Hv1 of intracellular organelles, being associated with the Warburg effect in early animal life. Recent observations with biogeochemical models (Shi *et al.*, 2022) allowed us to better visualize how the environment influenced the cellular homeostasis of the first animals and how these external conditions, among others, allowed to leverage changes in the cellular cytoskeleton and fusogenic properties that were determinant for the biophysical evolution of animals. Thus, ecological issues are extremely relevant and determinant of the beginning of animal life (Wood *et al.*, 2019). Cancer, as a factor to unravel metazoan evolution, is discussed in the works of Boveri (Boveri, 1929), who recognized that tumorigenesis is related to centrosome amplification. Ctenophores that were atentaculate (Parry *et al.*, 2021), polysperms, and at the base of metazoans (Ryan *et al.*, 2013; Moroz *et al.*, 2014; Arcila *et al.*, 2017; Shen, Hittinger and Rokas, 2017; Whelan *et al.*, 2017) pave the way to the recognition of our multiflagellate history linked to cytasters, with multiflagellate fusion at the origin of the first embryo.

From a current perspective, appropriation of embryology by physics is inevitable. Neoplasia and embryology, in the future, may open a route to sketch the genesis of the immune system, the evolution of ctenophore body forms, and the polyphyletic evolution of sponges. The close link of embryogenesis with cancer may help change our approach to animal models, allowing us to consider common properties that are very conserved in evolution, among them Neoplasia, as drivers of evolution. It is also believed that a multidisciplinary approach to evolution may alter the approach to cancer, as some of the answers to cure the disease may lie in the emergent properties of metazoans and in the real understanding of the physical and environmental context that allowed the formation of the first embryo and animal phylogeny.

**Acknowledgments**

I would like to express my sincere thanks to professors Juan Fernández and Viviana Cantillana, who, despite my short time with them in the laboratory of the Faculty of Sciences, University of Chile, inspired me, 25 years later, to hypothesize on the emergence of the first embryo, animal phylogeny, and cancer. I thank Dr. Jose Bastos, pathologist and oncologist, for his permanent contribution to my cancer research projects. The authors offer apologies to all researchers that could not be mentioned in the article, given the need to establish some priorities in the article's construction.

**Conflict of Interest Statement**

<p>The authors declare that the research was conducted in the absence of any commercial or financial relationships that could be construed as a potential conflict of interest.</p>

**Figures**

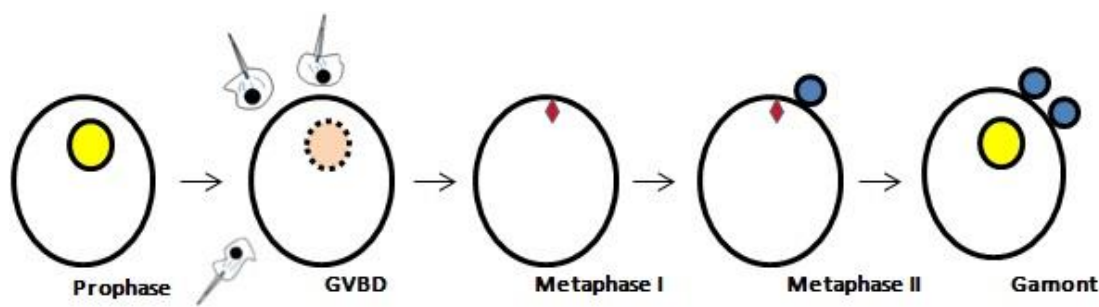

Figure 1. Fusion of multiflagellate microgametes with a meiotic intermediate. The fusion of various microgametes incorporated numerous cytasters in early animal evolution. Nuclear dimorphism could explain why only one of the nuclei fuses with the gamont nucleus. Most of the fused nuclei of microgametes become pycnotic. Participation of lamins and F-actin in cancer and embryogenesis suggests that the multiflagellate fusion event occurred before germinal vesicle breakdown.

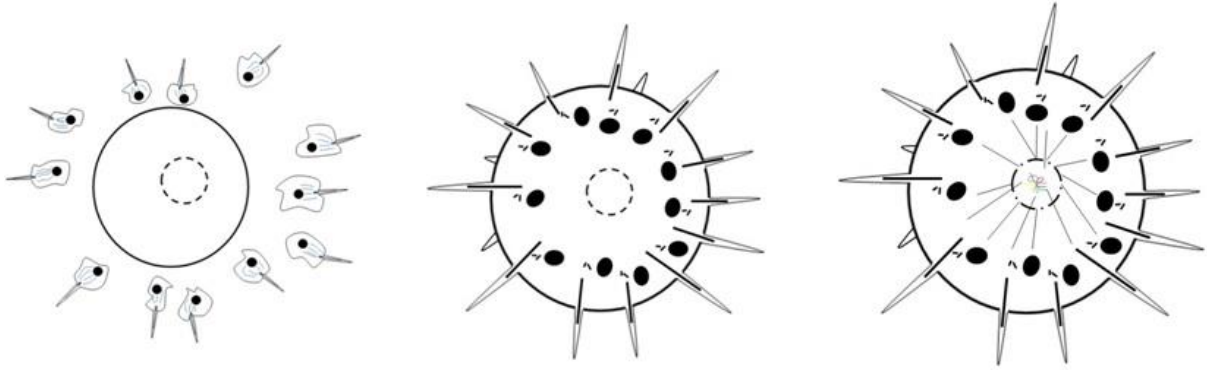

Figure 2. Model of multiflagellate fusion in early animal evolution. The initial event was eminently sensory, given the function of flagella in protists. Fusion records are found in cytasters embedded in the first zygote of metazoan evolution. Multiflagellate fusion was able to initiate the biophysical impregnation of the first embryo, having a relevant role in chromatin organization in this first nucleus. The mechanical memory incorporated in the first zygote allowed to reproduce the process after fusion of sperm cells and oocytes produced by the first hermaphrodite animal. In our view, the first embryo, cancer, and animal evolution emerged together.